\documentclass{article}    % Specifies the document style.
\usepackage{graphicx}
\title{The Role of Nonlocality in the Pinning Properties of Borocarbides Materials}  % Declares the document's title.
\author{A. V. Silhanek\footnote{Laboratorium voor Vaste-Stoffysica en Magnetisme, K. U. Leuven, Celestijnenlaan 200 D, B-3001 Leuven, Belgium.}, J. R. Thompson\footnote{Oak Ridge National Laboratory, Oak Ridge, Tennessee 37831-6061 and Department of Physics, University of Tennessee, Knoxville, Tennessee 37996-1200.}, L. Civale\footnote{Superconductivity Technology Center, MS K763, Los Alamos National Laboratory, Los Alamos, NM 87545, USA}, S. L. Bud'ko\footnote{Ames Laboratory, U.S. DoE and Department of Physics and Astronomy, Ames, Iowa, USA}, and P. Canfield\footnote{Ames Laboratory, U.S. DoE and Department of Physics and Astronomy, Ames, Iowa, USA}}         % Declares the author's name.

\begin{document}           % End of preamble and beginning of text.

\maketitle                 % Produces the title.

An experimental review on the influence of nonlocal electrodynamics in the vortex pinning properties of non-magnetic borocarbide superconductors is presented. We show that the pinning force density $F_p$ exhibits a rich and complex anisotropic behavior that sharply contrast with the small mass anisotropy of these compounds. For magnetic fields $\bf H$ applied parallel to the crystallographic c-axis, the first order reorientation transition between two rhombic lattices manifests itself as a kink in $F_p(H)$. For $\bf H \bot$c-axis, a much larger $F_p(H)$ and a slower relaxation rate is observed. In this field configuration, nonlocality induces a fourfold periodicity in $F_p$ when $\bf H$ is rotated within the square basal plane. Unlike the out-of-plane anisotropy, which persists for increasing impurity levels, the in-plane fourfold anisotropy can be strongly suppressed by reducing the electronic mean free path. This result unambiguously demonstrate that the in-plane anisotropy is a consequence of nonlocal effects.

%%%%%%%%%%%%%%%%%%%%%%%%%%%%%%%%%%%%%%%%%%%%%%%%%%%%%%%%%%%%%%%%%%%%%%%%%%%%%%%%%%%%%%%%%%%%%%%%%%%%%%%%%%%%%%%%%%%%%%%%%%
\section{INTRODUCTION}
%%%%%%%%%%%%%%%%%%%%%%%%%%%%%%%%%%%%%%%%%%%%%%%%%%%%%%%%%%%%%%%%%%%%%%%%%%%%%%%%%%%%%%%%%%%%%%%%%%%%%%%%%%%%%%%%%%%%%%%%%%

%MAGNETIC BOROCARBIDES%
The large attention that the borocarbide family of intermetallics (\mbox{RNi$_2$B$_2$C}, where R = rare earth) has received during the last years is in part due to the wide variety of interesting phenomena that these materials exhibit. Special attention has been devoted to R = Tm, Er, Ho and Dy, where superconductivity and antiferromagnetism coexist in a large portion of the phase diagram. The high superconducting transition temperatures $T_c$ and the broad variation of the ratio $T_N/T_c$ (where $T_N$ is the Neel temperature, ranging from $1.5$ K to $10$ K) make this family particularly appropriate to explore that coexistence\cite{yaron96,Eisaki94,Eskildsen98,Gammel99,Norgaard00}.

%NON-MAGNETIC BOROCARBIDES%
%NON-LOCALITY%
Even in the simplest case of non-magnetic borocarbides (R = Y, Lu), many remarkable properties have been reported. For example, in the superconducting mixed state the vortex lattice symmetry exhibits a strong field dependence with structural phase transitions separating a variety of exotic rhombic and square flux line lattices (FLL)\cite{yethiraj97,kogan97a,eskildsen97b,dewilde97,yethiraj98,mckpaul98,gammel99,sakata00}. In addition, the equilibrium magnetization $M$ deviates from the local London prediction\cite{song99} and exhibits a fourfold anisotropy when the field is rotated within the basal plane of these tetragonal materials\cite{civale99,kogan99}.

The presence of these non-hexagonal lattices has been attributed to the effects of nonlocal electrodynamics, which arise when the electronic mean free path $\ell$ is larger than the BCS zero temperature superconducting coherence length $\xi_0$. Nonlocal electrodynamics in superconductors had been traditionally associated with very low values of the Ginzburg-Landau parameter, $\kappa \sim 1$. Borocarbide superconductors have $\kappa$ values in the range of 10 to 20, which allows one to avoid cumbersome core treatments by using the very simple London approach and simultaneously introduce the nonlocal effects as a perturbative term. 

%DEVIATION OF Mrev FROM LOG BEHAVIOR%
This approach was first developed by Kogan et al.\cite{kogan96} to analyze the case of a superconductor with an isotropic gap and an isotropic cylindrical Fermi surface. In this scenario, the authors successfully explained the observed deviations of the reversible magnetization $M(H)$ of Bi:2212 in the mixed state from the logarithmic dependence, $M \propto ln(H_{c2}/H)$, predicted by the London model\cite{kogan88a}. Later on, Song et al.\cite{song99} found similar deviations in YNi$_2$B$_2$C when ${\bf H} \parallel c$-axis, that could also be quantitatively accounted for by this model. More recently we have extended that study to all field orientations and showed that this generalization of the London theory provides a satisfactory complete description of the anisotropic $M(H)$ with a self-consistent set of parameters\cite{basalplane}.

%GAP OR FERMI SURFACE?
In general, the additional nonlocal corrections introduced in the London equation gives rise to a coupling of the microscopic supercurrents in a vortex to the underlying crystal symmetry. As a consequence, nonlocality induces an anisotropic response even in cubic materials\cite{christen}, which according to the local London theory, should behave isotropically.

How actually the superconducting currents are connected to the crystal symmetry is still a matter of debate. First, Kogan et al.\cite{kogan97a,kogan96} developed a model considering a superconductor with an anisotropic Fermi surface and an order parameter with s-wave symmetry. Later on, Franz and co-workers\cite{franz} analyzed the opposite case of an isotropic cylindrical Fermi surface in a d-wave superconductor. The applicability of each one of these models depends eventually on the material studied.

In the particular case of non-magnetic borocarbides compounds, there is growing evidence that both an anisotropic Fermi surface\cite{dugdale,martinez} and a non conventional pairing mechanism\cite{martinez,kohara,jacobs,zheng,nohara,yokoya,yang,boaknin,izawa} are present. In the last years, the lack of any corroborative evidence for a non conventional pairing mechanism in these materials meant that most of the experimental results were interpreted within the Kogan et al.'s model. However, the proven existence of a pronounced anisotropy in the gap symmetry challenges the original view and strongly encourages one to critically re-examine several previous interpretations.

As we pointed out above, none of the introduced models has to be necessarily invoked in order to explain the deviation in the reversible magnetization from $M \propto ln(H_{c2}/H)$, whereas both of them correctly account for the anisotropic in-plane magnetization. However, still some differences appear between these approaches, when analyzing the field and temperature evolution of the vortex lattice.

%STRUCTURAL TRANSITIONS%
The model of Kogan et al. predicts\cite{kogan97a} that two structural transitions in the FLL should occur in borocarbides for ${\bf H} \parallel c$: a first order reorientation transition between two rhombic lattices at a field $H_1$ and a second order transition from rhombic-to-square at $H_2 > H_1$. According to this model, both transition fields shift up as temperature is increased. Recent small angle neutron scattering (SANS) studies\cite{mckpaul98,eskildsen01,mckpaul02} confirmed the existence of these transitions in YNi$_2$B$_2$C and LuNi$_2$B$_2$C. A jump in the apical angle $\beta$ of the rhombic lattice, discontinuous within the resolution, occurs at $H_1 \sim 1$ to $1.5$ kOe, and the lattice becomes square ($\beta = 90^{\circ}$) at $H_2 \sim 2$ to $2.5$ kOe\cite{mckpaul02}.

Franz et al.\cite{franz} showed that similar FLL rearrangements should appear in the case of a d-wave superconductor. However, two distinctive predictions emerge from this model. First, an additional first order transition should take place at very low temperatures. Second, in sharp contrast to the Kogan et al.'s model, the $H_1$ boundary should decrease as temperature is increased. In a recent work, Nakai et al.\cite{nakai} showed that when both anisotropic gap and Fermi surface are considered simultaneously, the intrinsic competition between these two effects leads to a much richer $H-T$ phase diagram.

\bigskip

%IRREVERSIBILITIES%
Although the role of nonlocality on the $\it {equilibrium}$ properties of the FLL has been convincingly established, much less attention has been paid to the effects on the dynamic or $\it {nonequilibrium}$ vortex response. The very low critical current density $J_c$ observed in non magnetic borocarbides for ${\bf H} \parallel c$, associated with large pinning correlation volumes\cite{eskildsen97b}, indicates that the elastic properties of the FLL must play a key role in the pinning. Since the shear modulus $C_{66}$ depends on the vortex lattice symmetry\cite{rosenstein99}, and the apex angle $\beta$ undergoes a discontinuous jump at $H_1$, it is expected that $C_{66}$ and hence the pinning properties change abruptly at this field\cite{eskildsen97b}. In other words, vortex pinning, which involves distortions from equilibrium vortex configurations, should be affected by the symmetry changes in the vortex lattice. Indeed, as we will show throughout this work, nonlocal effects influence the vortex pinning in several ways.

%IN THIS WORK
First, in both non-magnetic compounds YNi$_2$B$_2$C and LuNi$_2$B$_2$C, and for ${\bf H}\parallel c$, the reorientation transition at $H_1$ induces a kink in the pinning force density $F_p(H)$. We find that $H_1(T)$ slightly decreases as $T$ increases, in contrast to $H_2(T)$. We also study the effect of Co-doping in Lu(Ni$_{1-x}$Co$_x$)$_2$B$_2$C. We observe that $H_1$ decreases as the nonlocal effects progressively fade out when $x$ increases. These results are in agreement with the expected evolution of $H_1$ within a non conventional superconducting gap scenario.

Second, we observe a fourfold oscillation in $F_p(H)$ when ${\bf H}$ is rotated within the $ab$-plane (in-plane anisotropy). We show that this effect is strongly suppressed as the amount of impurities is increased, i.e. as the nonlocal effects are reduced. This clearly demonstrate that the in-plane anisotropy is a consequence of nonlocal effects.

Strikingly, we also observe a large anisotropy in $F_p$ between the c-axis and the basal plane (out-of-plane anisotropy). We found that $F_p$ for ${\bf H} \bot c$ is about one order of magnitude larger than for ${\bf H}\parallel c$ and has a quite different field dependence. Accordingly, a much faster normalized relaxation rate $S$ is detected for ${\bf H}\parallel c$. This effect persists for all the explored impurity concentrations, demonstrating that the out-of-plane anisotropy is not related with nonlocal effects. We also rule out the presence of surface barriers for ${\bf H}\bot c$ by performing minor hysterisis loops.

This work is organized as follows. In the next section we briefly introduce the experimental details and the samples studied. In section 3 we discuss the reorientational structural transition and how it influences the dynamic response. In section 4 we focus on the in-plane anisotropy of the pinning force density and in section 5 we discuss the out-of-plane anisotropy. Finally, in section 6 we state our conclusions.

%%%%%%%%%%%%%%%%%%%%%%%%%%%%%%%%%%%%%%%%%%%%%%%%%%%%%%%%%%%%%%%%%%%%%%%%%%%%%%%%%%%%%%%%%%%%%%%%%%%%%%%%%%%%%%%%%%%%%%%%%%
\section{EXPERIMENTAL ASPECTS}
%%%%%%%%%%%%%%%%%%%%%%%%%%%%%%%%%%%%%%%%%%%%%%%%%%%%%%%%%%%%%%%%%%%%%%%%%%%%%%%%%%%%%%%%%%%%%%%%%%%%%%%%%%%%%%%%%%%%%%%%%%

The composition, dimensions, critical temperature and estimated electronic mean free path for each of the studied single crystals is summarized in Table I. Normal state magnetization measurements in the Y-0 sample reflect the presence of a very diluted distribution of localized magnetic moments. The paramagnetic signal follows a Curie law which corresponds to a rare-earth impurity content of $0.1$ at. $\%$ relative to yttrium, probably due to contaminants in the yttrium starting material\cite{basalplane}. Similar measurements in the Lu-x single crystals show a much weaker Curie tail at low temperatures, $T < 100$ K, which might be due to a $0.001 \%$ magnetic impurities of Gd in the Lu crystallographic site\cite{cheon98}.

\begin{table*}[ht]
\begin{center}
\caption[]{sample name, composition, volume, thickness, superconducting critical temperature and electronic mean free path for the investigated samples.}
\begin{tabular}{llllll}                                    
\hline
sample & \multicolumn{1}{c}{composition} & \multicolumn{1}{l}{$V($mm$^3)$} & \multicolumn{1}{l}{$t($mm$)$} & \multicolumn{1}{l}{$T_c($K$)$} & \multicolumn{1}{c}{$l(\AA)$} \\ 
\hline
\hline
Y-0 & YNi$_2$B$_2$C & $2.8$ & $0.5$ & $15.1$ & $300$ \\
Lu-0 & LuNi$_2$B$_2$C & $2.5$ & $0.3$ & $15.7$ & $270$ \\
Lu-1.5 & Lu(Ni$_{1-x}$Co$_x$)$_2$B$_2$C (x=1.5$\%$) & $1.2$ & $0.4$ & $14.9$ & $100$ \\
Lu-3 & Lu(Ni$_{1-x}$Co$_x$)$_2$B$_2$C (x=3$\%$) & $0.44$ & $0.2$ & $14.1$ & $70$ \\
\hline
\end{tabular}
\end{center}
\end{table*}

Magnetization measurements were conducted on commercial Quantum Design SQUID magnetometers. Part of the relaxation measurements were taken in a MPMS-7 ($H \leq 7$ T) and the angular studies were performed in a MPMS-5S ($H \leq 5$ T) equipped with a home-made rotating system\cite{evidence,anomalous}. In the superconducting mixed state, isothermal magnetization $M(H)$ loops were recorded. For the angular studies, both components of ${\bf M}(H)$ were measured using the two sets of pick up coils (longitudinal and transverse). The critical current density $J_c$ was determined from the width of the hysteresis loops according to the Bean critical state model, as previously described\cite{evidence}. Relaxation measurements $M(t)$ were performed on both branches (increasing and decreasing) of the hysteresis loop over periods of 1 hour.

%%%%%%%%%%%%%%%%%%%%%%%%%%%%%%%%%%%%%%%%%%%%%%%%%%%%%%%%%%%%%%%%%%%%%%%%%%%%%%%%%%%%%%%%%%%%%%%%%%%%%%%%%%%%%%%%%%%%%%%%%%
\section{REORIENTATIONAL TRANSITION}
%%%%%%%%%%%%%%%%%%%%%%%%%%%%%%%%%%%%%%%%%%%%%%%%%%%%%%%%%%%%%%%%%%%%%%%%%%%%%%%%%%%%%%%%%%%%%%%%%%%%%%%%%%%%%%%%%%%%%%%%%%

According to the existing models,\cite{kogan97a,franz} two types of structural transitions in the vortex lattice should occur in borocarbides materials, namely a first order reorientation transition between two rhombic lattices at a field $H_1$ and a second order transition from the rhombic to a square lattice at a higher field $H_2$. These predictions were recently confirmed by small angle neutron scattering (SANS) experiments\cite{mckpaul98,eskildsen01,mckpaul02} 
{\it{on a sister crystal to the Y-0 sample studied here}}. For ${\bf H} \parallel c$ and at $T=4.5$ K, a jump in the apical angle $\beta$ of the rhombic vortex lattice occurs at $H_1 \sim 1$ to $1.5$ kOe, and the lattice becomes square ($\beta = 90^{\circ}$) at $H_2 \sim 2$ to $2.5$ kOe. 

\begin{figure}[hhh]
\centering
\includegraphics[angle=0,width=90mm]{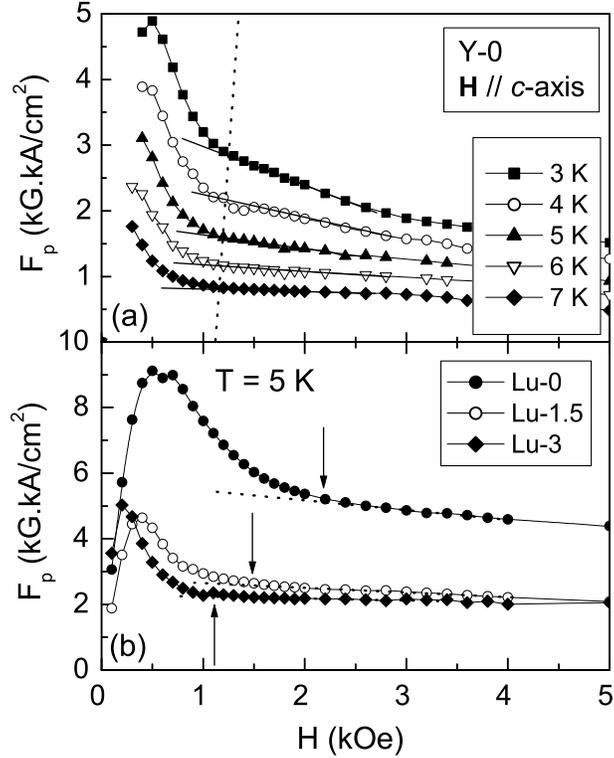}
\caption[]{{\small Field dependence of the pinning force density for $\bf H \parallel c$-axis for (a) Y-0 sample at several temperatures and (b) Lu-0, Lu-1.5 and Lu-3 samples at $T=5$ K.}}
\label{FpvsHc}
\end{figure}

As we pointed out above, it is expected that the discontinuous jump of the apex angle $\beta$ at $H_1$ induces an abrupt change in the pinning properties through the shear modulus $C_{66}(\beta)$\cite{eskildsen97b,rosenstein99}. In order to check this prediction, we measured the pinning force density $F_p=\left| \bf {J_c \times B} \right|$ for the \mbox{Y-0} sample as a function of the applied field for $\bf H \parallel c$-axis at several $T$ (see Figure \ref{FpvsHc}(a)). 

\begin{figure}[htb]
\centering
\includegraphics[angle=0,width=90mm]{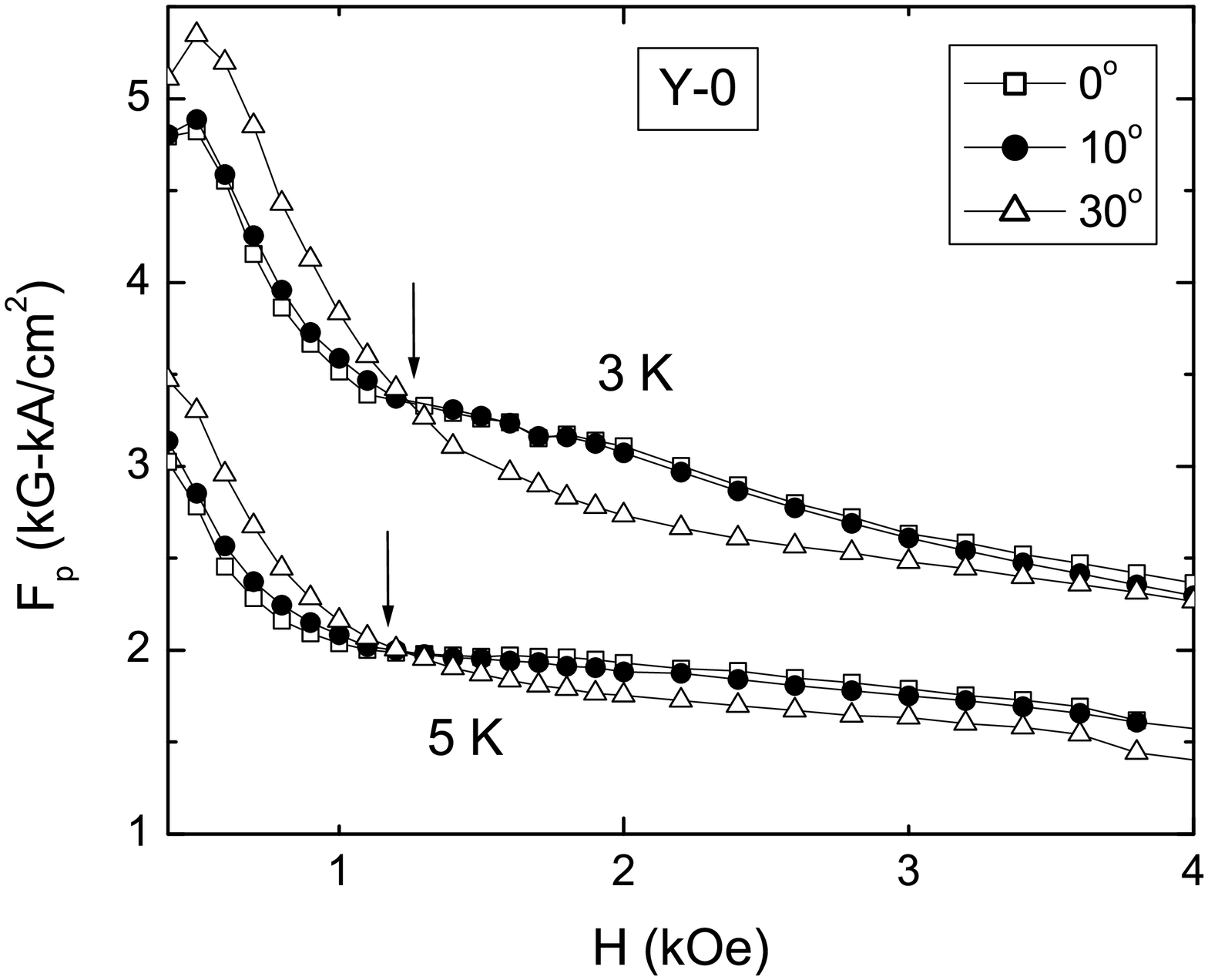}
\caption[]{{\small Field dependence of the pinning force density for three different angles of $\bf H$ with respect to the c-axis at $T=3$ and $5$ K.}}
\label{kinkvsangle}
\end{figure}

We observe that at low fields $F_p(H)$ strongly decreases with increasing $H$, up to a certain field $H^* \sim 1.2$ kOe, above which  the field dependence is much less pronounced. Several facts indicate that this jump in $dF_p/dH$ is a signature of the reorientation phase transition $H_1$ in the FLL\cite{Fp}. First, the position of the kink for $\bf H \parallel c$ coincides with the value of $H_1$ reported in the literature\cite{mckpaul02}. Second, as shown in  Figure \ref{kinkvsangle}, where $F_p(H)$ for sample Y-0 is plotted at several field orientations, $H^*$ is rather insensitive to the field orientation, in agreement with the behavior of $H_1$ determined from SANS experiments\cite{mckpaul98}.  

\begin{figure}[htb]
\centering
\includegraphics[angle=0,width=90mm]{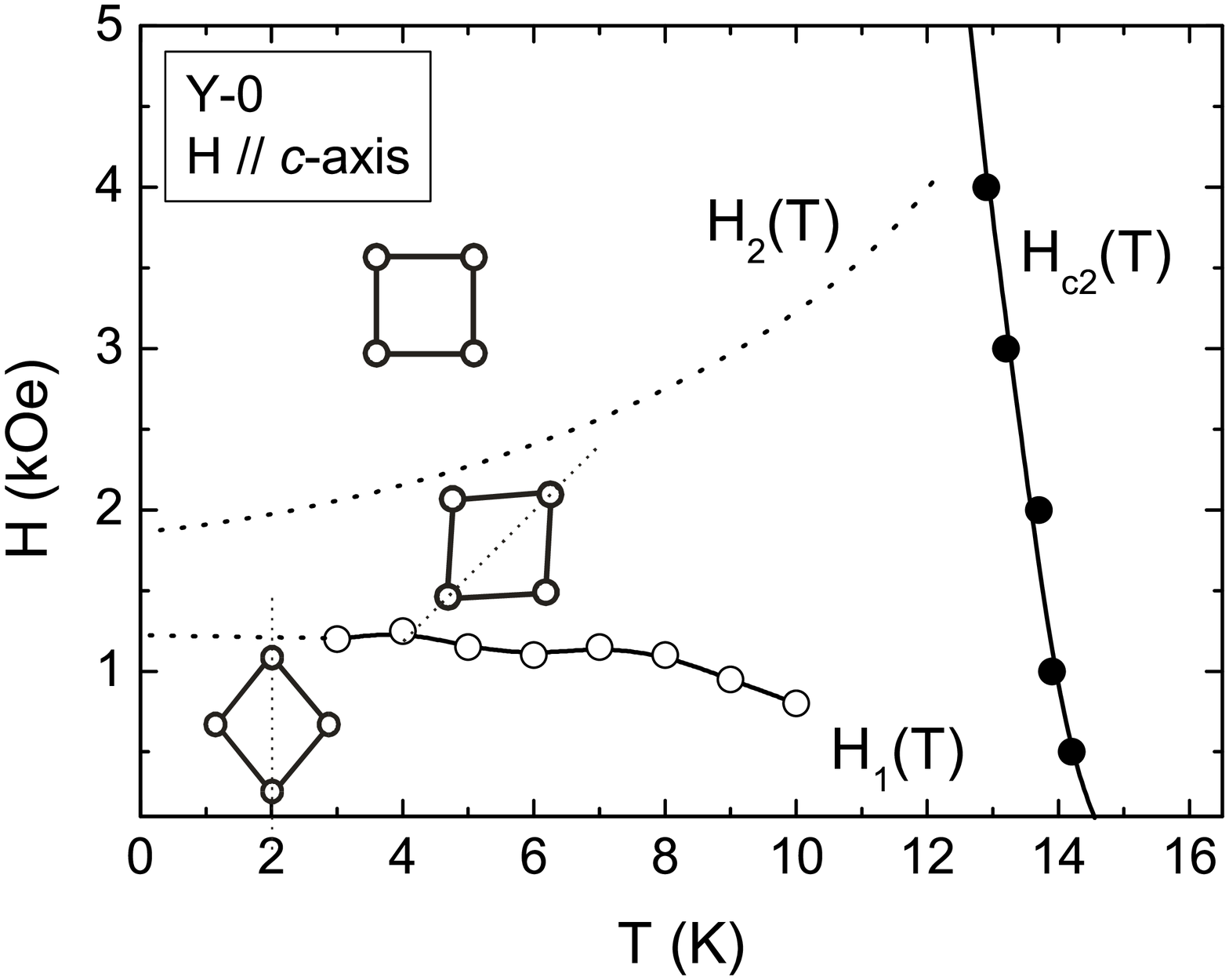}
\caption[]{{\small Temperature dependence of the upper critical field and the structural reorientation transition field for the Y-0 sample when ${\bf H}\parallel c$.}} 
\label{phasediagram}
\end{figure}

In Figure \ref{FpvsHc}(a), it can be also observed that the transition field $H^*(T)$ remains almost constant at low $T$ and, unlike the rhombic-to-square second order transition $H_2(T)$, it slightly decreases with $T$ at higher temperatures. This behavior is in agreement with recent neutron scattering experiments\cite{mckpaul02} and with theoretical predictions for a d-wave superconducting gap\cite{franz}. In contrast, it follows from the model of Kogan et al. that both transitions, $H_1$ and $H_2$, should {\it{increase}} as nonlocality vanishes. In Figure \ref{phasediagram} we show the temperature evolution of $H^*$ together with the upper critical field $H_{c2}$.

Further evidence that the observed feature at the field $H^*$ arises from nonlocal effects come from the dependency of this transition field with the electronic mean free path. Indeed, the strength of the nonlocal perturbations is parametrized by a new characteristic distance, the nonlocality radius $\rho(T,\ell)$: smaller $\rho(T,\ell)$ corresponds to weaker nonlocality effects.  This means that $\rho(T,\ell)$ decreases with increasing $T$ or decreasing $\ell$. Thus, if the position of the kink is related with nonlocal effects, an increase in $T$ or a decrease in $\ell$ should produce qualitatively similar effects\cite{kogan97a,kogan96}.

It has been shown that the Co substitutes on the Ni site in borocarbides compounds as a nonmagnetic impurity\cite{cheon98} that suppresses $T_c$\cite{schmidt}, but does not modify significantly the critical current $J_c$\cite{Gammel99}. This Co doping procedure allows one to accurately control the electronic mean free path by tuning the amount $x$ of impurities.

To determine the $\ell$ dependence of $H^*$, we performed measurements on the Lu-0, Lu-1.5 and Lu-3 samples for $\bf H \parallel c$ (estimates of $\ell$ are given in Table I). The results are shown in Figure \ref{FpvsHc}(b), where we plotted $F_p(H)$ at $T = 5$ K. We observe that the kink in $F_p$ (indicated by the arrows) is still visible in the field range where the transition $H_1$ should appear, and that it shifts to lower fields with increasing $x$, in agreement with the $T$ dependence. In other words, the analogy between $x$ and $T$ is satisfied for $H^*$, thus confirming that the kink arises from nonlocality. 

From Figure \ref{FpvsHc}(b) we also confirmed the fact that the introduction of Co does not affect the pinning properties. Another interesting fact is that the Lu-0 sample has a larger $F_p$ than the Y-0 sample at the same $T$, even though it has a lower density of magnetic impurities. This indicates that the magnetic moments in the Y-0 are not relevant pinning centers for the flux lines when ${\bf H}\parallel c$.

%%%%%%%%%%%%%%%%%%%%%%%%%%%%%%%%%%%%%%%%%%%%%%%%%%%%%%%%%%%%%%%%%%%%%%%%%%%%%%%%%%%%%%%%%%%%%%%%%%%%%%%%%%%%%%%%%%%%%%%%%%
\section{IN-PLANE ANISOTROPY}
%%%%%%%%%%%%%%%%%%%%%%%%%%%%%%%%%%%%%%%%%%%%%%%%%%%%%%%%%%%%%%%%%%%%%%%%%%%%%%%%%%%%%%%%%%%%%%%%%%%%%%%%%%%%%%%%%%%%%%%%%%

\begin{figure}[htb]
\centering
\includegraphics[angle=0,width=90mm]{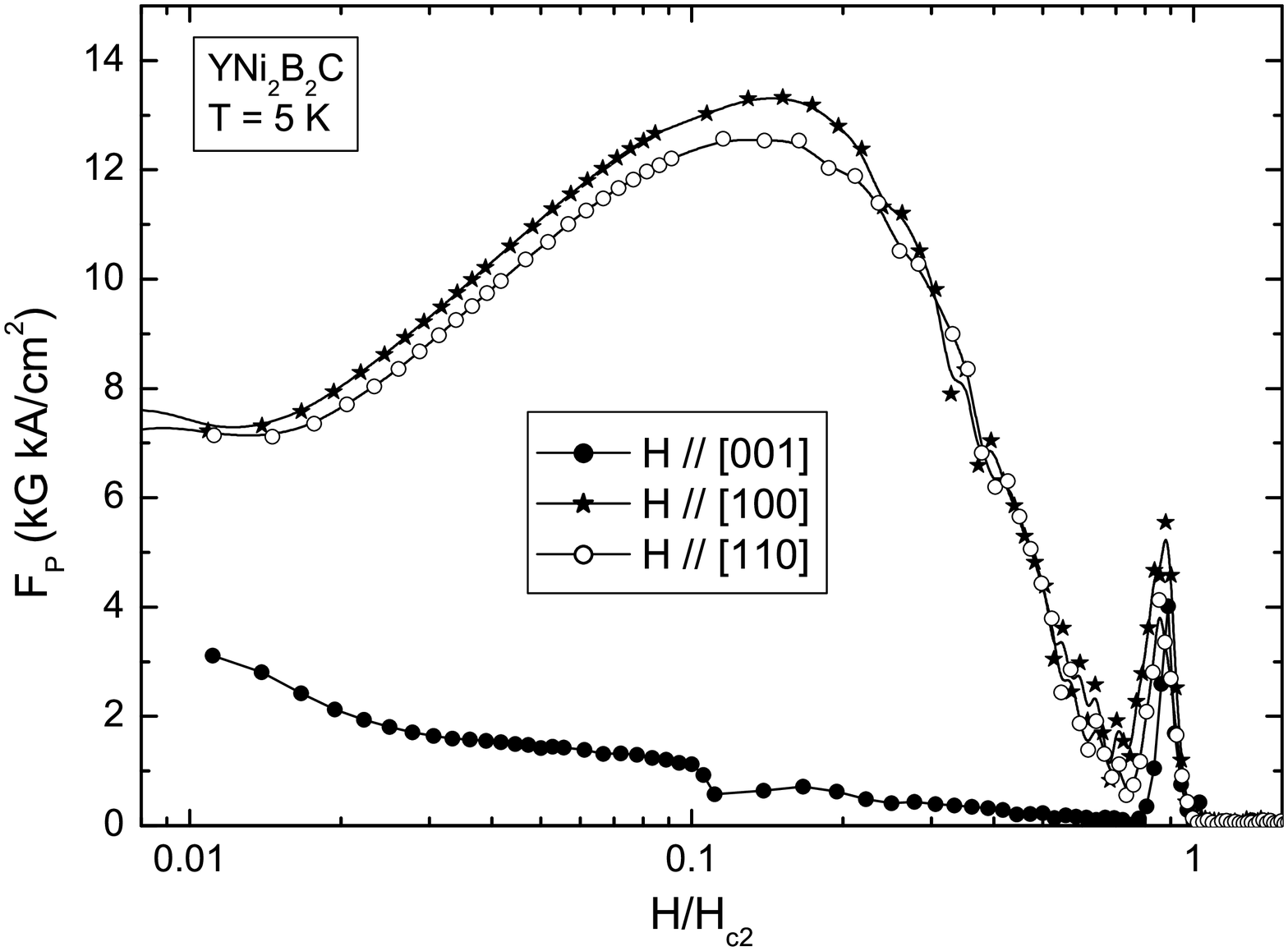}
\caption[]{{\small Pinning force density $F_p$ vs $H$  with $\bf H$ applied along the principal axes $[100]$, $[110]$ and $[001]$ at $T =$ 5 K.}} 
\label{FpvsHalldirections}
\end{figure}

Let us now analyze the pinning properties for ${\bf H} \bot c$. Figure \ref{FpvsHalldirections} shows $F_p(H)$ for ${\bf H}$ applied along the symmetry axes $[001], [100]$ and $[110]$, at $T=5$ K for the Y-0 sample. We observe that for the in-plane orientations, $F_p(H)$ exhibits a very different behavior than for ${\bf H} \parallel c$. First, over most of the field range $F_p({\bf H} \bot c) \gg F_p({\bf H} \parallel c)$. Second, the pinning force density exhibits a clear and systematic in-plane anisotropy, $F_p[100]>F_p[110]$. We now focus on the origin of this in-plane anisotropy and defer the discussion of the large out-of-plane anisotropy to the next section.

\begin{figure}[htb]
\centering
\includegraphics[angle=0,width=90mm]{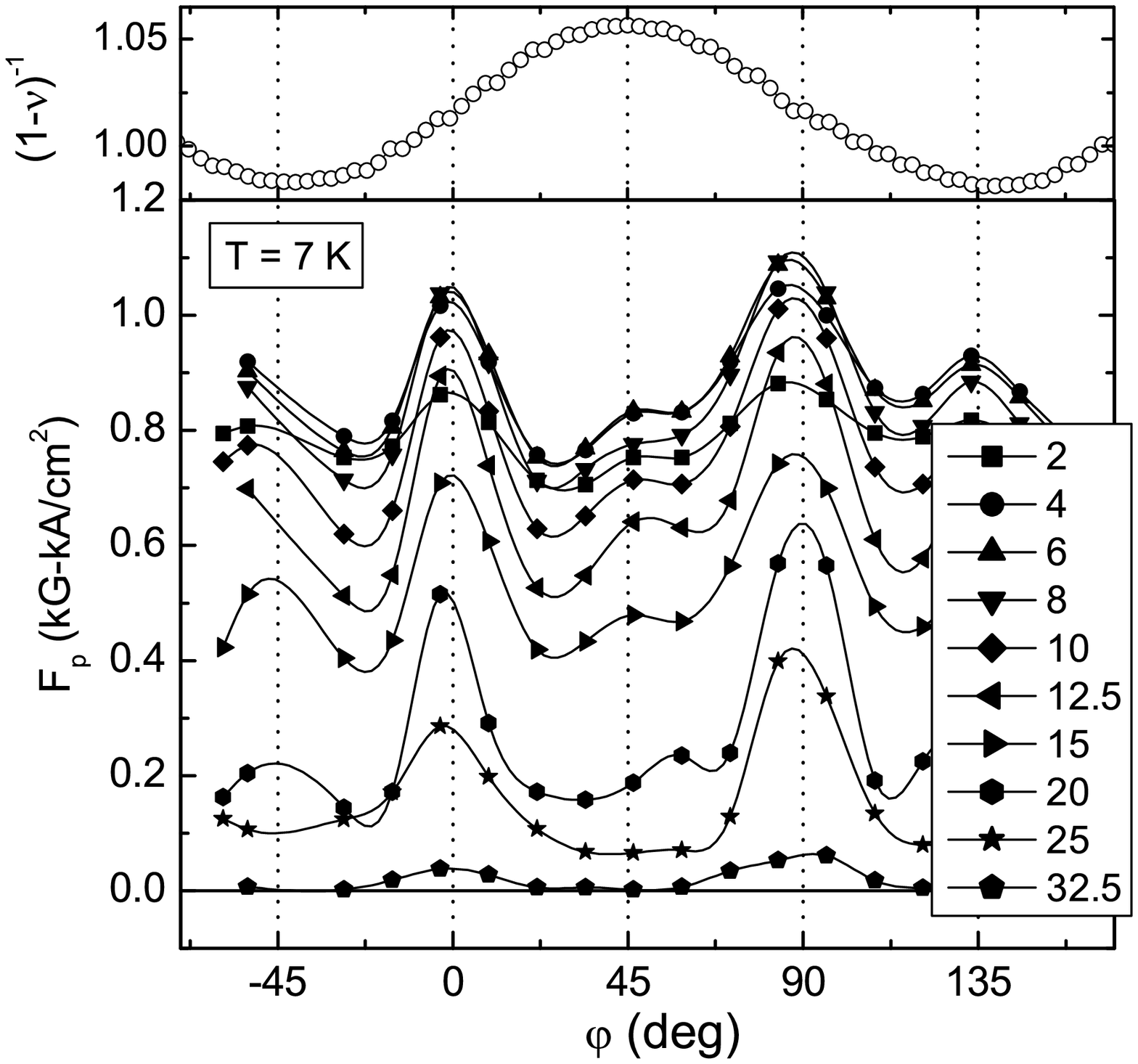}
\caption[]{{\small In-plane angular dependence of (a) Meissner slope $-4\pi dM/dH = (1- \nu)^{-1}$ at $T =$ 7 K (upper panel) and (b) pinning force density $F_p(\varphi)$ at $T=$ 7 K and several fields (lower panel), for the Y-0 single crystal. The labels indicate the applied field in kOe.}} 
\label{inplaneY0}
\end{figure}

The fourfold nature of the in-plane anisotropy becomes evident in the lower panel of Figure \ref{inplaneY0}, where we show the angular dependence of the pinning force density, $F_p(\varphi)$, at $T=7$ K for several $H$. This figure shows that for all $H>2$ kOe, $F_p(\varphi)$ exhibits a rather complex behavior: in addition to the main peaks at $[100]$ and $[010]$, secondary maxima are visible at $[110]$ and $[1\bar{1}0]$. 

\begin{figure}[htb]
\centering
\includegraphics[angle=0,width=90mm]{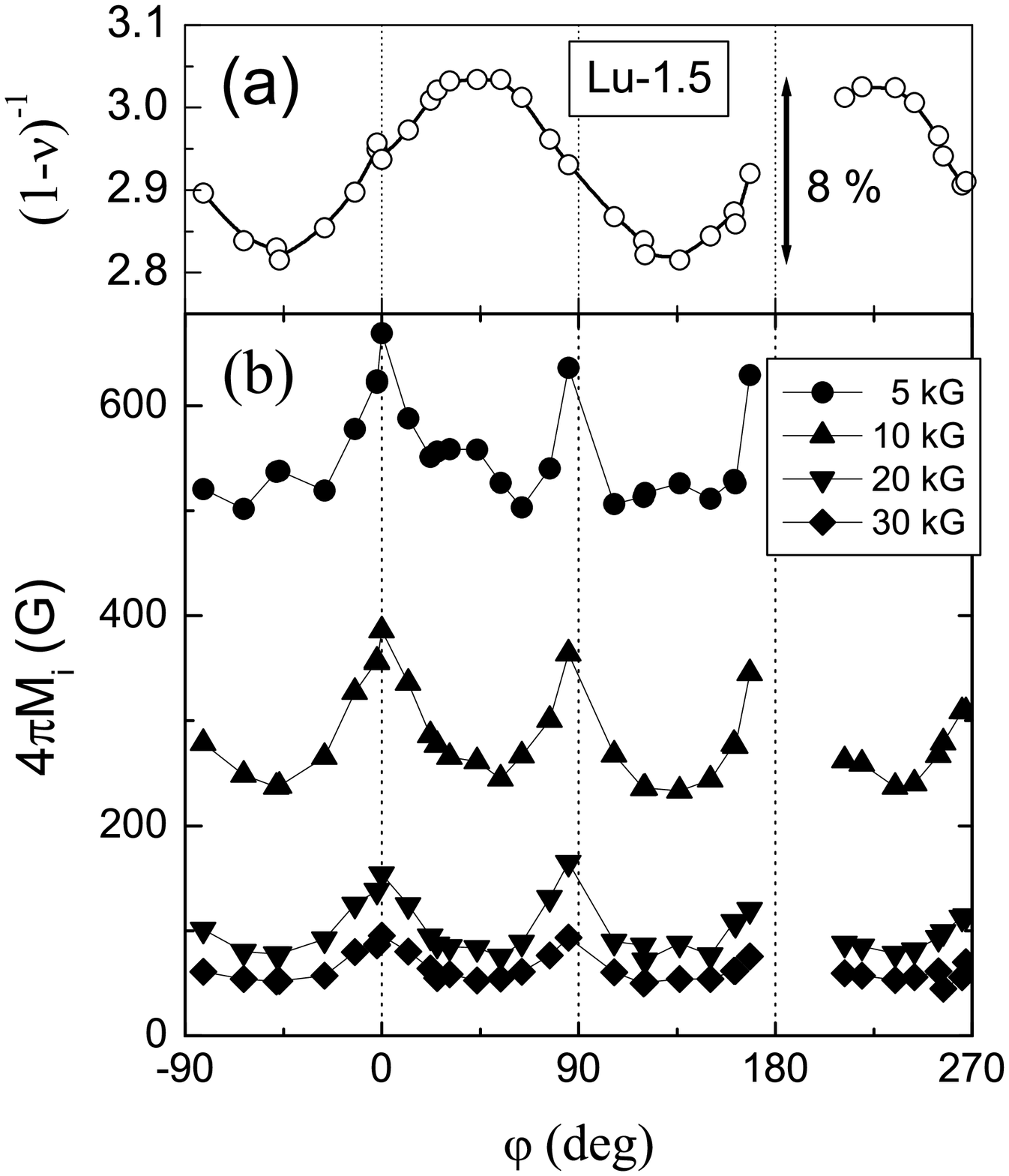}
\caption[]{{\small In-plane angular dependence of (a) Meissner slope $-4\pi dM/dH = (1- \nu)^{-1}$ at $T =$ 5 K (upper panel) and (b) irreversible magnetization $M_i(\varphi)$ at $T=$ 5 K and several fields (lower panel), for the Lu-1.5 single crystal.}} 
\label{inplaneLu15}
\end{figure}

\begin{figure}[htb]
\centering
\includegraphics[angle=0,width=90mm]{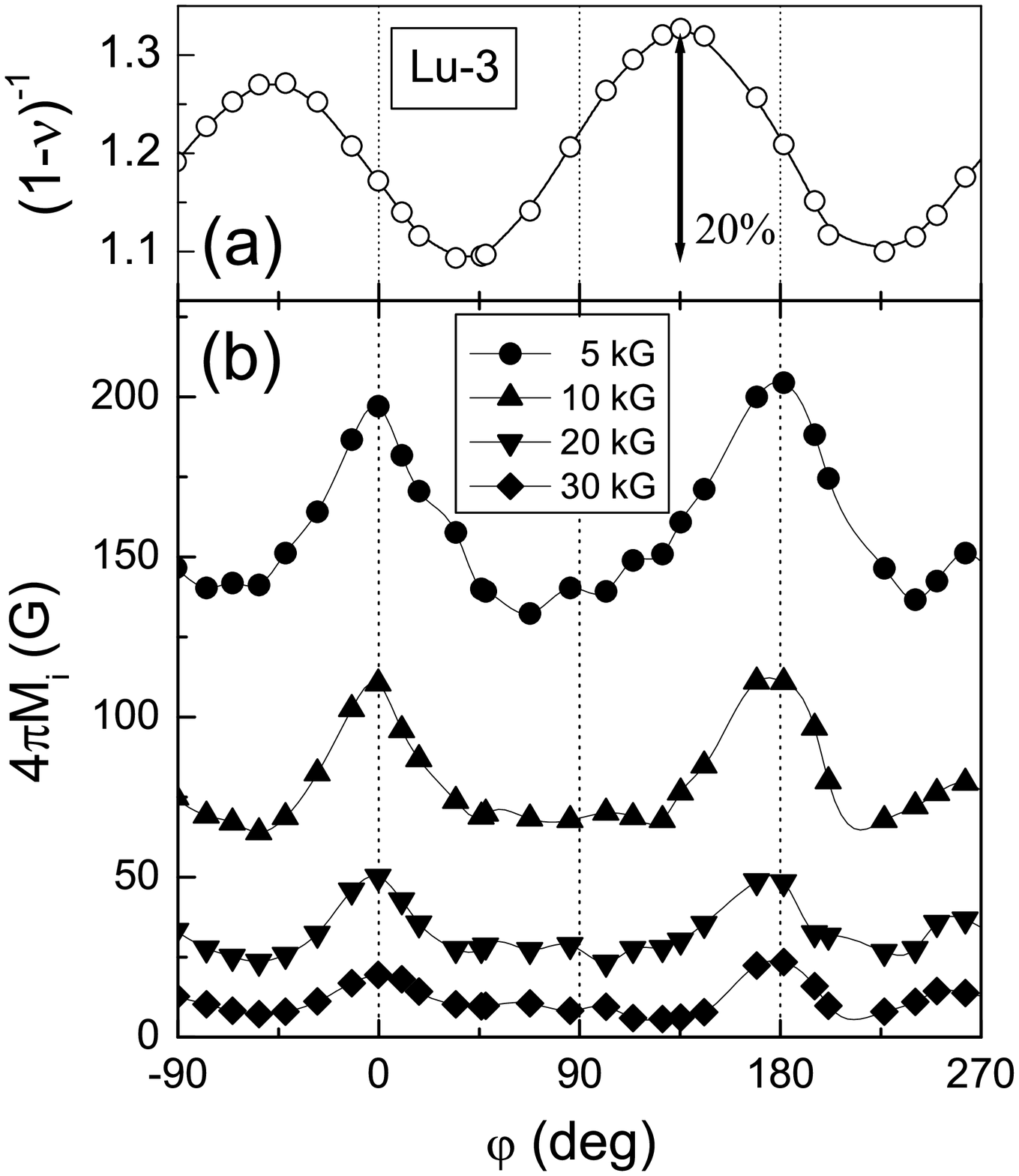}
\caption[]{{\small In-plane angular dependence of (a) Meissner slope $-4\pi dM/dH = (1- \nu)^{-1}$ at $T =$ 5 K (upper panel) and (b) irreversible magnetization $M_i(\varphi)$ at $T=$ 5 K and several fields (lower panel), for the Lu-3 single crystal.}} 
\label{inplaneLu3}
\end{figure}

For comparison, in the upper panel we show the in-plane angular dependence of the Meissner slope $-4\pi (dM/dH)$ for $H=30$ Oe at $T=$ 7 K. As $H$ = 30 Oe is well below the lower critical field $H_{c1}$ for all $\varphi$, this curve represents the total flux exclusion of the Meissner state. The oscillatory behavior with periodicity $\pi$ (two-fold symmetry) originates from purely geometrical effects. Indeed, a field applied at any orientation within the basal plane can be decomposed in $H_x = H cos(\varphi+45^{\circ})$ and $H_y = H sin(\varphi+45^{\circ})$. If we approximate the crystal shape by the ellipse, the Meissner response associated with each component is $4\pi M_i=-H_i/(1-\nu_i)$, where $i=x;y$ and $\nu_i=t/L_i$ are the demagnetizing factors, thus $4\pi M = -H [cos^2(\varphi+45^{\circ})/(1-\nu_x)+sin^2(\varphi+45^{\circ})/(1-\nu_y) ]$.
 
While it is very unlikely that any type of crystallographic defects could account for the complex in-plane variations in $F_p$, nonlocal effects can provide a natural explanation for them. Due to nonlocality, the geometry of the vortex lattice depends on the orientation within the plane. We can again argue (as in the ${\bf H} \parallel c$ case) that $C_{66}$ depends on the lattice geometry, and that such dependence must be reflected in $F_p$. Interestingly, $F_p$ has local maxima at the high symmetry crystallographic orientations $[100]$ and $[110]$, where $C_{66}$ is expected to exhibit local minima. In agreement with a nonlocal scenario, we have also observed that the in-plane anisotropy progessively decreases as $T$ approaches $T_c$\cite{Fp}.

Conclusive evidence that the in-plane fourfold anisotropy arises from nonlocal effects is provided by similar measurements conducted in the doped samples Lu-1.5 and Lu-3. As we pointed out above, by tuning the impurity content in the sample, we are able to control the electronic mean free path and so the nonlocal effects. Figure \ref{inplaneLu15}(b) shows the irreversible magnetization $M_i$ (proportional to the pinning force density) as a function of $\varphi$ for several fields at $T=5$ K. Even though in this sample $\ell$ is about three times smaller than in the Y-0 crystal, the fourfold anisotropy is still observed. For comparison, in Fig. \ref{inplaneLu15}(a) we show the Meissner slope $-4\pi(dM/dH)=(1-\nu)^{-1}$ which reflects the sample geometry effects. The smallness of the amplitude of this twofold oscillation indicates that the sample shape is closer to a disk.

Interestingly, further decreasing the mean free path down to $\ell=70 \AA$ by increasing the impurity concentration level 
to $x=3\%$, converts the oscillatory response from a fourfold to a twofold anisotropy (see Figure \ref{inplaneLu3}). Surprisingly, the irreversible magnetization does not maximize at the same angular position as the Meissner curve, but there is a shift of nearly 45$^\circ$ between them. This is an unexpected misalignment that cannot be accounted for by a bianisotropic current distribution, but may be related to the triangular shape of the crystal (viewed along the c-axis).

%%%%%%%%%%%%%%%%%%%%%%%%%%%%%%%%%%%%%%%%%%%%%%%%%%%%%%%%%%%%%%%%%%%%%%%%%%%%%%%%%%%%%%%%%%%%%%%%%%%%%%%%%%%%%%%%%%%%%%%%%%
\section{OUT-OF-PLANE ANISOTROPY}
%%%%%%%%%%%%%%%%%%%%%%%%%%%%%%%%%%%%%%%%%%%%%%%%%%%%%%%%%%%%%%%%%%%%%%%%%%%%%%%%%%%%%%%%%%%%%%%%%%%%%%%%%%%%%%%%%%%%%%%%%%

In the previous section we showed that the pinning force density in the Y-0 sample is much larger for ${\bf H} \bot c$ than for ${\bf H} \parallel c$. This out-of-plane $F_p$ anisotropy sharply contrasts with the very small ($< 10 \%$) mass anisotropy\cite{civale99,kogan99,basalplane}, therefore, explanations based on the anisotropic scaling frequently used in high $T_c$ superconductors\cite{blatter92,interplay} can be ruled out.

\begin{figure}[htb]
\centering
\includegraphics[angle=0,width=90mm]{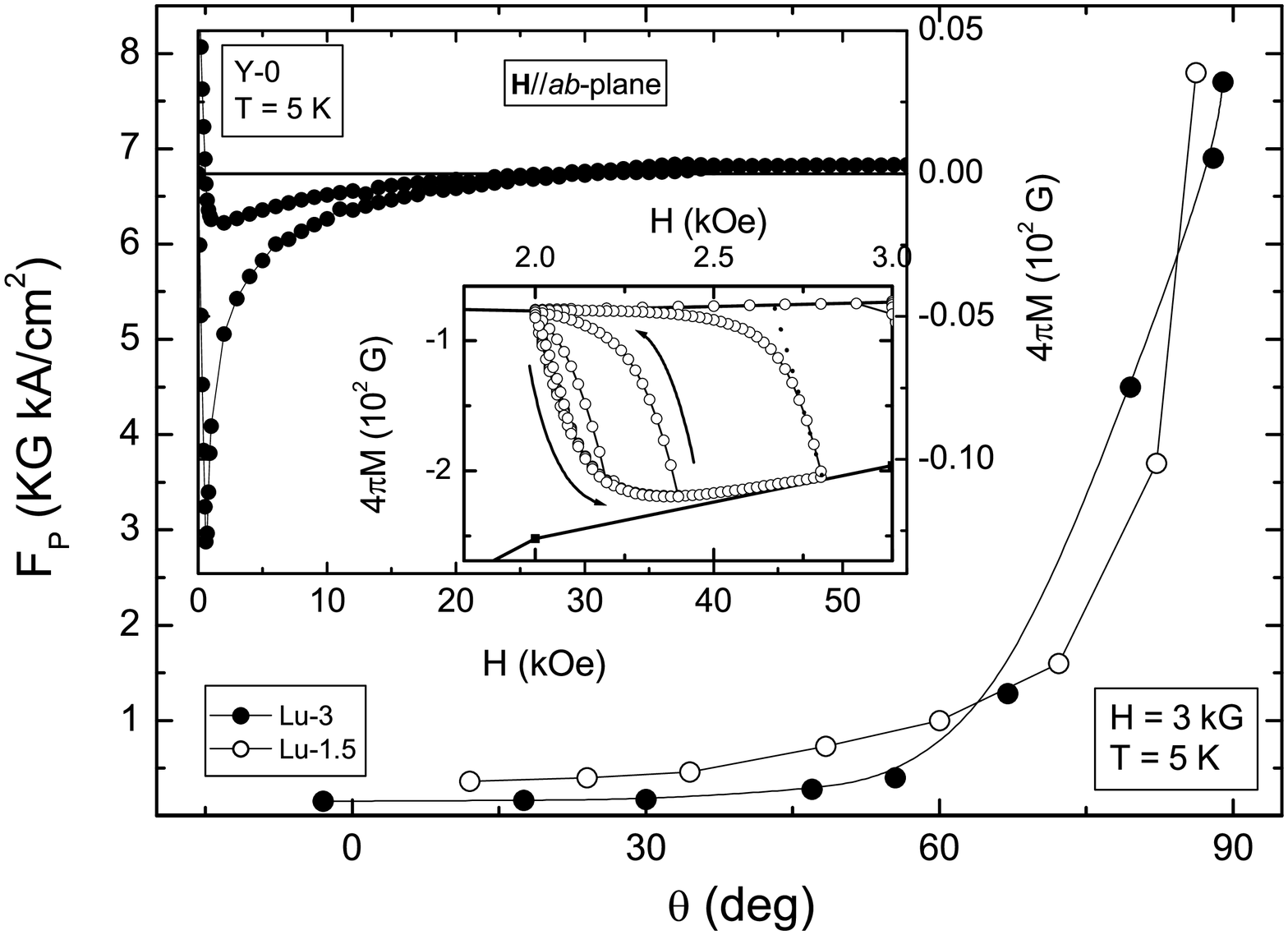}
\caption[]{{\small Main panel: pinning force density $F_p$ as a function of the angle $\theta$ between $\bf H$ and the c-axis, for the Lu-1.5 and Lu-3 samples at $T = 5$ K and $H = 3$ kG. The outer inset shows an hysterisis loop for ${\bf H}\bot c$ for the Y-0 sample at $T = 5$ K. A zoom in of the minor hysterisis loops corresponding to the full loop showed in the outer inset is shown in the inner inset. The dotted line shows the behavior expected for surface barriers.}} 
\label{minorloops}
\end{figure}

\begin{figure}[htb]
\centering
\includegraphics[angle=0,width=100mm]{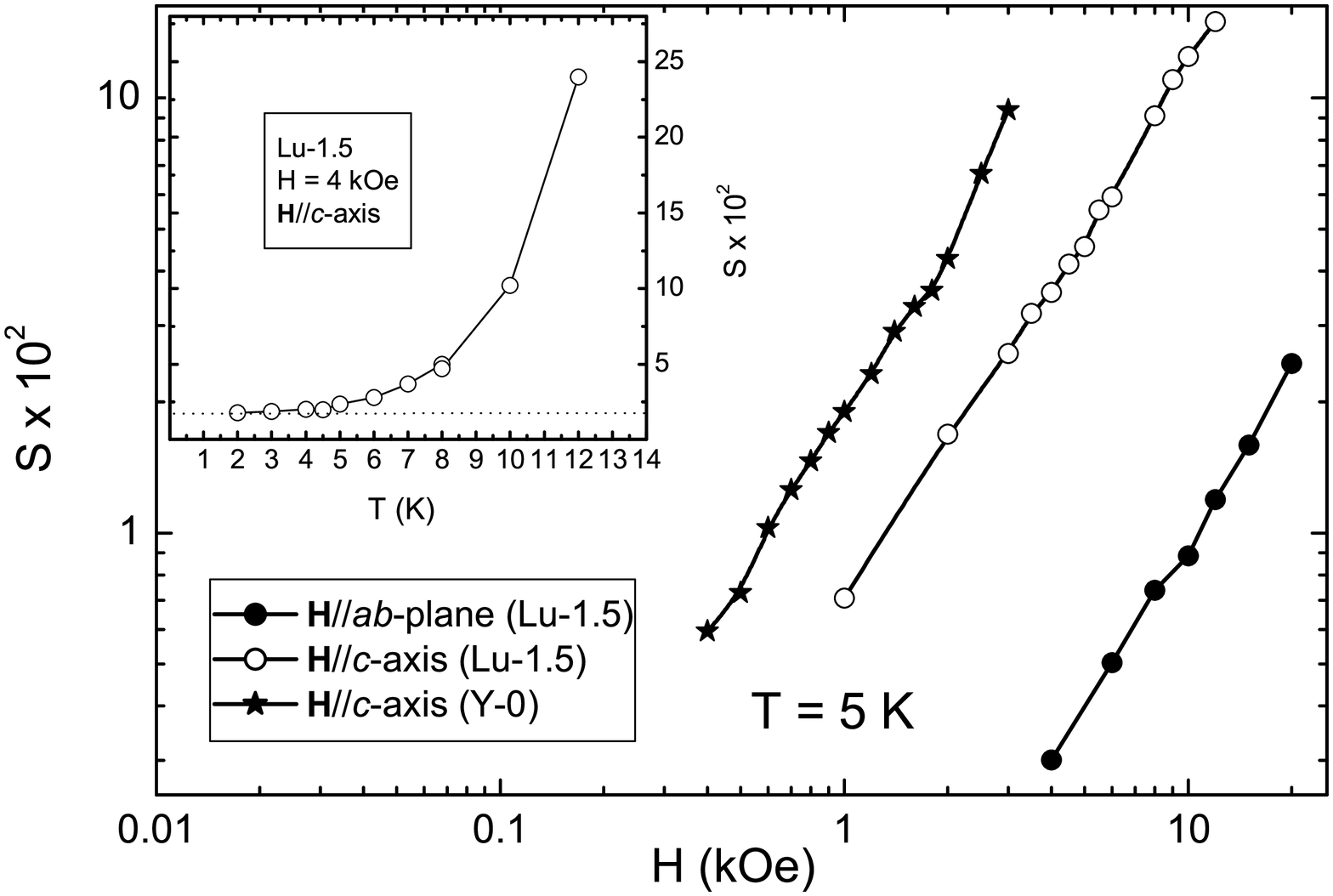}
\caption[]{{\small Main panel: normalized time relaxation rate $S(H)$ for different samples and field orientations. Inset: $S$ vs. $T$ for the Lu-1.5 sample at $H = 4$ KOe and ${\bf H} \parallel c$.}} 
\label{creep}
\end{figure}

%NON LOCALITY%
In the same way, this effect is too large to be ascribed to nonlocality, which should appear as a perturbatively small effect. This conclusion is confirmed by the presence of similar anisotropy values in the doped samples Lu-1.5 and Lu-3, where nonlocal effects are strongly reduced. Indeed, in the main panel of Figure~\ref{minorloops} we show $F_p(\theta)$, where $\theta$ is the angle between the applied field and the c-axis, for the Lu-1.5 and Lu-3 at $T=5$ K and $H=3$ kOe. As the field is progressively tilted off from the c-axis, we first observe an almost angle independent pinning force $F_p$ up to $\theta \sim 60^\circ$. Beyond this angle, $F_p$ increases very fast up to nearly one order of magnitud larger than for $\theta=0^\circ$, when $\bf H$ appoaches the basal plane. As we have previously demonstrated, this out-of-plane anisotropy is field dependent and drops to almost isotropic behavior near the peak effect (see Fig.~\ref{FpvsHalldirections}).

%SURFACE BARRIERS%
One possible reason for the sharp increase of $F_p(\theta)$ as $\theta \rightarrow 90^\circ$ could be the presence of significant surface barriers for ${\bf H} \bot c$. One way to check whether the surface barriers play a relevant role in the measured pinning force is to perform minor hysteresis loops with ${\bf H} \parallel ab$ at several $T$ and $H$. In the larger inset of Figure~\ref{minorloops} we show an examples for the Y-0 sample at $T=$ 5 K. If surface barriers were the main source of hysteresis, no flux changes would occur in the bulk while $H$ is changing from one branch of the main loop to the other one, hence the data of the minor loop connecting the lower and upper branches would be Meissner-like straight lines\cite{ullmaier75}. In contrast, in the case of bulk pinning, the lines connecting both branches are curved (parabolic in the simplest Bean model for an infinite slab) just as we observe in the smaller inset of Figure \ref{minorloops}. Moreover, in a recent work\cite{ucrania} we have demonstrated that $F_p$ calculated from these minor loops assuming only bulk pinning are in good agreement with those obtained from the main loops. Thus, a significant contribution to magnetic hysteresis arising from surface barriers can be ruled out. Furhermore, it is worth noting that the peak effect (near $H_{c2}$) which is a bulk phenomena, is observed in the whole field orientation, clearly indicating that bulk pinning dominates the magnetic response\cite{xiao}.

%CREEP
A fact that we have not considered up to now is that, due to the significantly large time relaxation rates of the persistent currents in these materials, the current density $J$ determined through magnetization measurements in the typical time scale of the SQUID-magnetometers ($\sim 20$ sec) is smaller than the ``true'' critical current density $J_c$. This suggests that the observed large out-of-plane anisotropy may be related to an angular dependence of the time relaxation of $J$. 

%SvsH%
To explore this possibility, we have measured the normalized time relaxation rate of the irreversible magnetization $S = -d\ln J/d\ln t$ for the Y-0 and Lu-1.5 samples for ${\bf H} \parallel c$ and ${\bf H} \bot c$ at several fields. These measurements are shown in Figure~\ref{creep}. Surprisingly, in both samples the creep rate $S$ for ${\bf H} \parallel c$ is comparable or even larger than the values obtained in HTSC at the same temperature (5 K). In the simplest Anderson-Kim scenario, disregarding the possible influence of both quantum creep and glassy relaxation (see below), the pinning energy $U_p$ is given by $U_p \approx T/S$. For instance, in the particular case of the Lu-1.5 sample at $H = 6$ kOe and $T = 5$ K, we obtain $U_p \sim 85$ K, which is low even for HTSC standards. With regards to these values, it is worth keeping in mind that the scale of the {\it elementary} pinning energy $u_p$ is $\left( H_c^2/8\pi \right) \epsilon \xi^3$, where $H_c$ is the thermodynamic critical field and $\epsilon$ is the mass anisotropy. In the HTSC, the low pinning energy is associated to the small $u_p$, which in turn is due to the very small coherence length. In contrast, in the borocarbide crystals the small value of $U_p$ is a consequence of the very low density of defects.

%creep anisotropy%
A remarkable fact observed in Fig.~\ref{creep} is the large difference (almost one order of magnitude) of the creep rate between the two field orientations studied for the Lu-1.5 sample. We additionally find that for both orientations, and also for sample Y-0 (in the $H \parallel c$ orientation) $S \propto H^n$ with an exponent $n=1.28(2)$. Considering again the simplest scenario, $U_p$ and $F_p$ are related by $U_p \approx F_p V_c \xi$ where $V_c$ is the flux line lattice correlation volume. By combining this expression with $S \approx T/U_p$, we obtain $S \approx T/F_p V_c \xi$, so the creep rate anisotropy should be approximately the inverse of the $F_p$ anisotropy, as indeed observed. This result confirms that the source of the observed anisotropies is the anisotropic pinning energy $U_p$. So, to complete the solution of this issue an explanation for the anisotropy in $U_p$ must be found.

%[LEO: fijate si podes decir algo de esto. Este valor esta cerca al 1.33 que encontraron en MgB2(cond-mat/0104112)]. 

%PLANAR DEFECTS%
In brief, the origin of the large out-of-plane pinning anisotropy is unclear. A simple explanation for it could be the presence of some still unidentified anisotropic pinning centers, such as planar defects. Clearly, the complexity of the angular dependence of $F_p$ deserves further investigation.

%QUANTUM CREEP: SvsT%
Another puzzling result comes from the study of the temperature dependence of the creep rate $S(T)$ for fixed fields. In the inset of Figure~\ref{creep} we show $S(T)$, for the Lu-1.5 sample at $H = 4$ kOe and ${\bf H} \parallel c$. We observe that $S$ decreases monotonically as $T$ decreases and it seems to saturate at low temperatures ($T < 4$ K) in a value $S \sim 0.02$. Both the finite value of $S$ as temperature approaches to zero and its temperature independent behavior are typical indications of a quantum creep process. However, an estimation of normalized creep rate (near $j_c$) is given by $S_Q = \hbar/(\eta \xi^{2} L_c)$,\cite{blatter94} where $\eta$ is the friction coefficient for a vortex line and $L_c$ is the length of the tunneling vortex segment. This relation can be written as $S_Q \sim Q_u \sqrt{j_c/j_0}$ where $Q_u = e^2 \rho_n/ \hbar \epsilon \xi$, $\rho_n$ is normal-state resistivity and $\epsilon$ the mass anisotropy. For the Lu-1.5 sample, $\rho_n \approx 4.2 \times 10^{-6} \Omega$ cm, $\epsilon \sim 1.1$ and $j_c/j_0 \sim 10^{-5}$, therefore $S_Q \sim 4 \times 10^{-6}$, four orders of magnitude smaller than the measured value. The preceding analysis assumes that the dynamics of the quantum relaxation is dissipative. However, in our borocarbide crystals the electronic mean free path $\ell$ is considerably larger than $\xi_0$. This raises the possibility that the quantum relaxation may be more accurately described by a Hall-type vortex dynamics, which is supposed to dominate in the super-clean limit. In that case, an analogous expression for $S_Q$ should apply, with $\eta$ replaced by a Hall coefficient $\alpha \gg \eta$\cite{blatter94}. Thus, if that were the case, the numerical discrepancy with the experimental result would be even larger.
 
An alternative interpretation could be that the observed $T$-independent behavior corresponds to the well known plateau range predicted by the glassy relaxation and widely observed in HTSC. If that were the case, further reducing the temperature (to $T < 1$ K) should result in a decreasing creep rate that tends to zero as $T$ approaches zero. In this scenario, in the plateau range $S^{-1} \sim \mu \ln (1+t/t_0)$, and hence $S \sim 0.02$ implies that the glassy exponent $\mu \approx 2$ (assuming, as is usually done, that $\ln (1+t/t_0) \sim 25$). However, the $\mu$ value obtained from the Maley analysis\cite{maley} is only around 0.3. Moreover, a $\mu \approx 2$ should produce very clear deviations from logarithmic relaxation, which we have failed to observe. A natural way to define this issue would be to perform a set of long term relaxation measurements ($\sim$ 12 to 24 hours). Another problem with this interpretation is that it implies an extremely small pinning energy, $U_p \ll T \mu \ln (1+t/t_0)$ for $T > 2$ K, that is, $U_p \ll 100$ K.

In summary, although the observed $S(T)$ is suggestive of quantum creep, a quantitative description remains elusive at this point. The alternative explanation in terms of a glassy plateau also has serious difficulties. This complex situation highligths the relevance of the measurements of the normalized relaxation rate $S$. Indeed, $S$ is a fundamental variable of the dynamics of the vortex system, whereas the persistent current density $J$ is measured at a particular time, and thus it is a more derived variable that depends on both the ``initial" critical current $J_c$ and the relaxation rate.  
Clearly, more detailed studies of the flux creep phenomena in borocarbide superconductors are desirable.

As a final application of the creep results, let us use them to estimate the dimensions of the correlation volume. As mentioned above, at $H = 6$ kOe and $T = 5$ K, we have $U_p \sim 85$ K. At this field value $F_p \sim 2.12$ kG.kA/cm$^2$, which results in a correlation volume $V_c \approx 10^{10} \AA^3$. Because of the diluted distribution of pinning centers of these compounds, the Larkin-Ovchinnikov(LO) theory should apply without restrictions. According to this model $V_c = R_c^2 L_c$, where $R_c$ and $L_c$ are the dimensions of the vortex bundle perpendicular and parallel to the field direction and $L_c/R_c \sim \sqrt{2C_{44}/C_{66}}$, with $C_{44}$ the tilt modulus and $C_{66}$ the shear modulus of the FLL. Within this description $C_{44} \approx H^2/4\pi$ and in the simplest case of a triangular FLL in an isotropic medium with $\kappa \gg 1$ in the intermediate field regime $C_{66} \approx H_{c1}H(1-h^2)/16\pi$\cite{brandt77}. From this we can easily compute the size of the LO correlation length in the field direction $L_c \sim 10 R_c \sim 10^4 \AA$, which in units of the intervortex distance $a_0$ gives $L_c/a_0 \sim 20$ for $H = 6$ kOe. This value turns out to be one order of magnitude smaller than that obtained by SANS measurements\cite{eskildsen97b} at $T = 2.2$ K in Lu-0 and Y-0 crystals. However, we have naively assumed that $C_{66} \sim H_{c1}H/4\pi$ which according to the recent work of Knigavko et al.\cite{knigavko} is a overestimation of the total shear modulus. Indeed, in that work the authors demonstrate that the squash component $C_{66}$ decreases as $H$ increases, vanishes at the transition field $H_1$ and then grows slowly. In addition, Eskildsen et al.\cite{eskildsen97b} showed that for fields above $10$ kOe the size of the correlation volume is dominated by the properties of the shear modulus. Moreover, using the expression $F_p \propto C_{44}^{-1}C_{66}^{-2}$ corresponding to the LO scenario and the experimental $F_p(H)$, we computed $C_{66}(H)$, and  surprisingly abtain that $C_{66}$ decreases monotonically with $H$ both below and {\it above} $H_1$, in contradiction with the standard behavior\cite{Fp}. This reinforce the idea that $C_{66}$ values derived from LO theory are larger than the ``real'' shear modulus and therefore the obtained $L_c$ tends to be somewhat smaller than those observed by SANS experiments.

%%%%%%%%%%%%%%%%%%%%%%%%%%%%%%%%%%%%%%%%%%%%%%%%%%%%%%%%%%%%%%%%%%%%%%%%%%%%%%%%%%%%%%%%%%%%%%%%%%%%%%%%%%%%%%%%%%%%%%%%%%
\section{CONCLUSIONS}
%%%%%%%%%%%%%%%%%%%%%%%%%%%%%%%%%%%%%%%%%%%%%%%%%%%%%%%%%%%%%%%%%%%%%%%%%%%%%%%%%%%%%%%%%%%%%%%%%%%%%%%%%%%%%%%%%%%%%%%%%%

The high quality of the borocarbides single crystals have the double benefit of producing large mean free paths and a very dilute distribution of vortex pinning centers. The combination of these factors favours the observation of the influence of nonlocality on vortex pinning. In this work, we have shown that nonlocality significantly affects the irreversible properties of these materials.  With the magnetic field directed along the $c$-axis, the pinning force density $F_p(H)$ changes abruptly at the vortex reorientation field $H_1$, which is controlled by nonlocal effects.  In addition, its dependence on temperature and electron scattering (mean free path) further ties the observed structure in $F_p$ to nonlocal electrodynamics.  With the magnetic field in the square $ab$-basal plane, the four-fold periodicity in $F_p$ and $M_{irr}$ (which is reminiscent of the four-fold periodicity in the {\it equilibrium} magnetization) again shows a coupling of the vortex lattice to the crystalline lattice; this higher order coupling (greater than a second rank mass-anisotropy tensor) arises from nonlocal effects, which again are washed out by thermal or electron-scattering disorder.  These experimental observations demonstrate that nonlocality has a surprisingly profound influence on vortices, the flux line lattice, and 
vortex pinning in these remarkable materials.

However, the rather complex angular dependence of the pinning force density clearly suggests that other mechanisms besides nonlocality are involved as well. In particular, the origin of the large out-of-plane anisotropy remains unclear. We have conclusively ruled out explanations based on the mass anisotropy, pinning by magnetic impurities, non-local effects and surface barriers. On the other hand, we have observed that this effect is also present in the time relaxation rate $S$, thus confirming that the source of the observed out-of-plane anisotropy is the anisotropic pinning energy $U_p$. On top of this, the creep rate $S$ for ${\bf H} \parallel c$ exhibits two additional remarkable properties. First, the observed $S$ values are comparable or even larger than the values obtained in HTSC at the same temperature. Second, it is shown that $S$
saturate at low temperatures in a finite value, likely owing to a quantum creep process. These unexpected behaviors cannot be quantitatively accounted for by the usual creep models for HTSC. Clearly, further experimental and theoretical studies are necessary to gain insight into this rather unexplored and fascinating field of the influence of nonlocality on the
vortex pinning.

%%%%%%%%%%%%%%%%%%%%%%%%%%%%%%%%%%%%%%%%%%%%%%%%%%%%%%%%%%%%%%%%%%%%%%%%%%%%%%%%%%%%%%%%%%%%%%%%%%%%%%%%%%%%%%%%%%%%%%%%%%
\section{ACKNOWLEDGMENTS}
%%%%%%%%%%%%%%%%%%%%%%%%%%%%%%%%%%%%%%%%%%%%%%%%%%%%%%%%%%%%%%%%%%%%%%%%%%%%%%%%%%%%%%%%%%%%%%%%%%%%%%%%%%%%%%%%%%%%%%%%%%

We are pleased to thank the Atomic Energy Commission of Argentina and Oak Ridge National Laboratory, where the measurements presented in this review have been performed. We also want to thank the CONICET of Argentina for financial support. The experimental results described in this review were obtained in collaboration with C.V. Tomy and D. McK. Paul. We acknowledge many valuable discussions with H.R. Kerchner, Hyun Jeong Kim, D.K. Christen, M. Yethiraj. A.V.S. is supported by the Fund for Scientific Research, Belgium. Oak Ridge National Laboratory is managed by UT-Batelle, LLC for the United States Department of Energy under contract No. DE-AC05-00OR22725.

%%%%%%%%%%%%%%%%%%%%%%%%%%%%%%%%%%%%%%%%%%%%%%%%%%%%%%%%%%%%%%%%%%%%%%%%%%%%%%%%%%%%%%%%%%%%%%%%%%%%%%%%%%%%%%%%%%%%%%%%%%

\end{document}